\documentclass[sigconf,natbib=true]{acmart}
\AtBeginDocument{%
  }

\setcopyright{acmlicensed}
\copyrightyear{2018}
\acmYear{2018}
\acmDOI{XXXXXXX.XXXXXXX}

\acmConference[SIGIR]{SIGIR}{June 03--05, 2018}{Woodstock, NY}

\usepackage[table]{xcolor}
\usepackage{multirow}
\usepackage{xspace}
\usepackage{balance}
\usepackage{enumitem}

\begin{document}

\title{A Picture of Agentic Search}

\author{Francesca Pezzuti}
\email{francesca.pezzuti@phd.unipi.it}
\orcid{0009-0005-2364-2043}
\affiliation{%
  \institution{University of Pisa}
  \city{Pisa}
  \country{Italy}
}

\author{Ophir Frieder}
\email{ophir@ir.cs.georgetown.edu}
\orcid{0000-0001-5076-8171}
\affiliation{%
  \institution{Georgetown University}
  \city{Washington}
  \state{DC}
  \country{USA}
}

\author{Fabrizio Silvestri}
\email{fsilvestri@diag.uniroma1.it}
\orcid{0000-0001-7669-9055}
\affiliation{%
  \institution{Sapienza University of Rome}
  \city{Rome}
  \country{Italy}
}

\author{Sean MacAvaney}
\email{sean.macavaney@glasgow.ac.uk}
\orcid{0000-0002-8914-2659}
\affiliation{%
  \institution{University of Glasgow}
  \city{Glasgow}
  \country{UK}
}

\author{Nicola Tonellotto}
\email{nicola.tonellotto@unipi.it}
\orcid{0000-0002-7427-1001}
\affiliation{%
  \institution{University of Pisa}
  \city{Pisa}
  \country{Italy}
}


\newcommand{\mypar}[1]{\vspace{0.5em}\noindent\textit{{#1}.}}

\newcommand{\ourbenchmark}[0]{\texttt{ASQ}\xspace}
\newcommand{\autorefine}[0]{Autorefine\xspace}
\newcommand{\qwenmed}[0]{Qwen-7B\xspace}
\newcommand{\qwensmall}[0]{Qwen-3B\xspace}
\newcommand{\arun}[0]{arun\xspace}
\newcommand{\aruns}[0]{\arun{}s\xspace}
\newcommand{\hotpotqa}[0]{HQA\xspace}
\newcommand{\msmarcodev}[0]{MSM\xspace}
\newcommand{\researchyq}[0]{RQ\xspace}
\newcommand{\naturalq}[0]{NQ\xspace}
\newcommand{\agenttag}[1]{\small{\texttt{<#1>}}\xspace\normalsize}
\newcommand{\thinktag}[0]{\agenttag{think}}
\newcommand{\searchtag}[0]{\agenttag{search}}
\newcommand{\retrievedtag}[0]{\agenttag{information}}
\newcommand{\answertag}[0]{\agenttag{answer}}
\newcommand{\refinetag}[0]{\agenttag{refine}}
\newcommand{\datatype}[1]{{\small{\texttt{#1}}}}
\newcommand{\strtype}[0]{\datatype{str}}
\newcommand{\inttype}[0]{\datatype{int}}
\newcommand{\colname}[1]{{\small\texttt{#1}}}

\begin{abstract}
With automated systems increasingly issuing search queries alongside humans, Information Retrieval (IR) faces a major shift. Yet IR remains human-centred, with systems, evaluation metrics, user models, and datasets designed around human queries and behaviours. Consequently, IR operates under assumptions that no longer hold in practice, with changes to workload volumes, predictability, and querying behaviours.
This misalignment affects system performance and optimisation: caching may lose effectiveness, query pre-processing may add overhead without improving results, and standard metrics may mismeasure satisfaction.
Without adaptation, retrieval models risk satisfying neither humans, nor the emerging user segment of agents. However, datasets capturing agent search behaviour are lacking, which is a critical gap given IR's historical reliance on data-driven evaluation and optimisation. We develop a methodology for collecting all the data produced and consumed by agentic retrieval-augmented systems when answering queries, and we release the Agentic Search Queryset (\ourbenchmark) dataset. \ourbenchmark contains reasoning-induced queries, retrieved documents, and thoughts for queries in HotpotQA, Researchy Questions, and MS MARCO, for 3 diverse agents and 2 retrieval pipelines. The accompanying toolkit enables \ourbenchmark to be extended to new agents, retrievers, and datasets.

\vspace{0.6em}
\noindent\hspace{10em}\includegraphics[width=1.25em,height=1.25em]{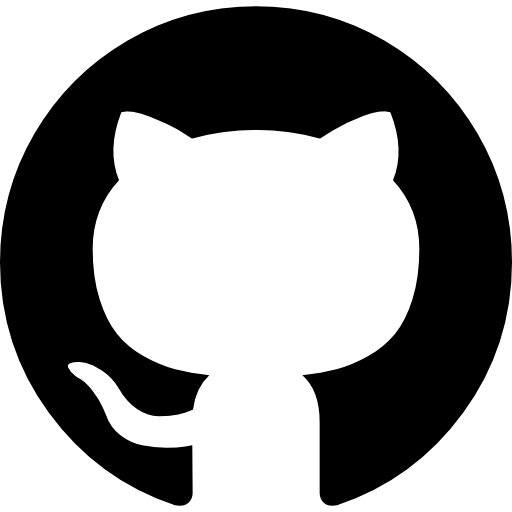}\hspace{.2em}
\parbox[c]{\columnwidth}{%
    \vspace{-.55em}\href{https://github.com/fpezzuti/ASQ}{\nolinkurl{fpezzuti/ASQ}}%
}

\vspace{0.1em}
\noindent\hspace{7em}\includegraphics[width=1.25em,height=1.25em]{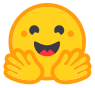}\hspace{.2em}
\parbox[c]{\columnwidth}{%
\vspace{-.55em}\href{https://huggingface.co/datasets/AgenticSearchQueryset/ASQ}{\nolinkurl{AgenticSearchQueryset/ASQ}}%
}
\vspace{-2.2em}

\end{abstract}

\ccsdesc[500]{Information systems~Information retrieval}

\keywords{Agentic Search, Information Retrieval, Evaluation Dataset}

\maketitle

\section{Introduction}

Information Retrieval (IR) has traditionally focused on optimising search systems for human users through rigorous, data-driven evaluation. In addition to the classic \textit{ad hoc} retrieval task, considerable attention focused on session-level search behaviours. In session-level search, systems model users' behaviour across a sequence of queries, often representing their attempts to fully address an information need~\cite{jansen2000ipm} or solve tasks~\cite{shah2023takingsearch}. Studies into session-level search yielded insights that  shaped the design and implementation of today's search engines, including system optimisation~\cite{zuo2022sessionsearch}, personalisation~\cite{zhang2025personalisation,macavaney2022personalisation}, and query suggestion~\cite{bacciu2024generatingqueryrecommendationsllms}. Behavioural data were collected from human users, including query logs~\cite{pass2006aol,zhang2006behaviour} and controlled examples from shared tasks~\cite{kanoulas2010trec,DBLP:journals/corr/abs-2003-13624, aliannejadi2024ikat}.

A major shift is happening. Increasingly, search engine queries are provided by automated systems. While a human user's query (or prompt) still usually initiates the process, Large Language Models (LLMs) are increasingly used to decompose the user's request and issue multiple queries to help fulfil it. 
This was initially popularised by the Retrieval-Augmented Generation (RAG) paradigm~\cite{gao2023retrieval,lewis2020rag}, which integrates IR modules into a cascade architecture where a retriever collects query-relevant documents from a document collection, and feeds them to the LLM generator that produces a final direct response.
This paradigm has since developed in terms of the technical search mechanism (from pre-generation retrieval~\cite{lewis2020rag} to general-purpose tool use~\cite{DBLP:conf/iclr/YaoZYDSN023,wu2025agentic}) and its applications (from simple question decomposition~\cite{rosset2025researchyquestions} to deep research~\cite{DBLP:journals/corr/abs-2112-09332}). These efforts are broadly referred to as \textit{agentic}~\cite{DBLP:journals/corr/abs-2410-09713}, where LLMs act as ``agents'' capable of taking actions autonomously to complete a task. Specifically, we focus on \textit{agentic RAG systems} (or \textit{reasoning-augmented search agents}) like Search-R1~\cite{jin2025searchr1} and AutoRefine~\cite{shi2025autorefine}, which extend RAG through an iterative approach to overcome its limitations in answering questions requiring multi-hop retrieval and reasoning~\cite{lewis2020rag, plaat2025surveyreasoning}.

With agents issuing queries alongside humans~\cite{zhai2025agenticretrieval,koneva2025internettraffic}, search systems are now exposed to two interleaved streams of queries: a fraction $\alpha$ of queries is human-generated (\textit{organic query stream}), the remaining fraction $(1-\alpha)$ of queries is issued by agents (\textit{synthetic query stream}). Notably, the fraction $\alpha$ of organic search queries\footnote{A separate problem relates to the prevalence of synthetic \textit{documents} (colloquially, the ``Dead Internet Theory''~\cite{walter2025artificial}). In this work, we instead focus on synthetic \textit{queries}.} is rapidly diminishing~\cite{zhai2025agenticretrieval}.
Since IR has historically been human-centred (implicitly assuming $\alpha=1$), with models, benchmarks, and evaluation methods designed around the information-seeking behaviours of human end-users, research has extensively characterised the organic stream as such.
By contrast, synthetic stream remained under-characterised and under-represented.

The two streams differ in several aspects.
Agents can generate queries at high speed and often produce reasoning-induced sub-queries~\cite{lin2025surveyagents, huang2025surveydeepresearch,plaat2025surveyreasoning}, leading to substantially higher query volumes.
Moreover, human-generated traffic typically follows seasonal and diurnal patterns~\cite{silvestri2010miningquerylogs,pass2006aol}, whereas machines-generated workload is less predictable~\cite{koneva2025internettraffic, wang2025burstgpt}.
Finally, there are differences in writing style between LLMs and humans~\cite{zendel2025llmqueries}, which can influence retrieval effectiveness~\cite{alaofi2022queryvariability}.
These differences have major implications for the design, optimisation, and evaluation of IR systems.  
For instance, recent work has argued well-established IR assumptions like the Probability Ranking Principle~\cite{robertson1977probability}, user models, benchmarks, evaluation methods, and even the notion of relevance may need to be revisited to serve agents~\cite{zhai2025agenticretrieval,tian2025ragrelevance,sauchuk2022role}.

Unfortunately, behavioural datasets, query logs, or benchmarks capturing how agents generate queries and retrieve relevant information do not yet exist. This gap is critical given their historical role in driving IR progress.
While synthetic \textit{queries}, generated using frameworks such as \textsc{QueryGym}~\cite{bigdeli2025querygym}, are already commonly used for query reformulation~\cite{ran2025llmgenqueryvariants}, user simulation~\cite{zendel2025llmqueries}, and data augmentation~\cite{askari2023trec2023}, they are produced artificially, often to resemble organic queries. Thus, they do not accurately reflect the information-seeking behaviours of AI agents.

Our contributions are as follows.
First, we develop a methodology to build datasets specifically designed to capture the search behaviours characterising the RAG agent user segment. We log their intermediate synthetic queries, retrieved documents, and reasoning descriptions they produce or consume by intercepting the retrieval calls during their decoding processes. Using this methodology, we release the \ourbenchmark dataset. 
 \ourbenchmark is based on HotpotQA (test)~\cite{yang2018hotpotqa}, Researchy Questions~\cite{rosset2025researchyquestions}, and MS MARCO dev~\cite{bajaj2016msmarco}, whose qrels are publicly available --- enabling both the optimisation and evaluation of retrieval methods. Since inference with agentic RAG systems is resource demanding and time consuming, \ourbenchmark's release contributes to sustainable and accessible research.
Due to the fast progress of agentic systems, we recognise that \ourbenchmark may become outdated. To address this, we also publicly release the code repository for extending \ourbenchmark with future agents.
We focus on open-source agents to facilitate reproducibility and ensure that \ourbenchmark can be used in online evaluations with the exact same agent, rather than commercial agents that are prone to undocumented updates and removal.

\section{Related Work}

\mypar{Query Sources}
IR has a long history of using realistic data to drive progress. Even as early as the Cranfield experiments, there have been concerns regarding the realism of search engine queries used in benchmarks~\cite{robertson2008ireval}. Since then, the community has often sourced organic queries from active search engines. For example, the TREC-8 Large Web Track sourced queries from Alta Vista, a popular commercial search engine at the time~\cite{hawking1999trecweb}. More recently, popular datasets such as MS MARCO~\cite{bajaj2016msmarco} and Natural Questions~\cite{kwiatkowski2019nq} have sourced their queries from Microsoft Bing and Google, respectively. On a few occasions, search engine logs that contain more detailed session information have been made available~\cite{pass2006aol,zhang2006behaviour}, though their use in research has been limited by ethical and legal considerations. Instead, datasets like the TREC Sessions Track~\cite{kanoulas2010trec}, TREC CAST~\cite{DBLP:journals/corr/abs-2003-13624}, and TREC iKAT~\cite{aliannejadi2024ikat} enable researchers to analyse session-level query behaviour, while UQV100~\cite{bailey2016uqv100} allows researchers to analyse retrieval robustness to reformulations (which may arise in search sessions).

These datasets support the evaluation and optimisation of retrievers across diverse query types and corpora. However, they are all reflective of the \textit{organic query stream}, limiting their suitability for agentic IR research.
To fill this gap, we develop a methodology to build datasets that capture how these agents search, and we release the \ourbenchmark dataset of agent-generated traces.

\mypar{Agentic RAG}
RAG systems generate responses conditioned on retrieved results~\cite{gao2023retrieval,lewis2020rag}, mitigating the well-known limitations of static and domain-specific parametric knowledge~\cite{lewis2020rag}. However, they struggle with complex or out-of-domain queries requiring multiple retrievals~\cite{lin2025surveyagents,plaat2025surveyreasoning}.
To address this, \textit{agentic RAG systems}, or \textit{agents} for short, such as Search-R1~\cite{jin2025searchr1}, AutoRefine~\cite{shi2025autorefine}, and related approaches \cite{jin2025searchr1,zhao2025parallelsearch} were proposed. These systems reframe question-answering as a sequential decision-making problem, implementing a closed feedback loop that interleaves chain-of-thought reasoning~\cite{wei2022cot}, retrieval, and generation~\cite{plaat2025surveyreasoning,wei2022cot}.

Optimised with reinforcement learning over a combination of factoid and multi-hop queries, \textit{agents} learn when and what to search. During inference, instead of answering after a single retrieval, they iteratively generate specialised XML-style control tags, to coordinate reasoning, retrieval, and generation within the so-called \textit{reasoning loop}.
During this loop, which terminates upon answer generation or encountering run-time errors, agents reason over retrieved documents, integrate the acquired knowledge into their context, and formulate new queries to address other eventual knowledge gaps.
This loop implements a multi-turn interaction between a generator and a retriever that allows agents to effectively solve simple QA tasks using parametric knowledge or a few retrieval steps, and to use multiple reasoning-conditioned searches to solve more complex ones.

Agent differences primarily stem from their state–action space and reward design, and thus, in their learned behaviours. For example, while Search-R1 state-action space only includes retrieval, answer and query generation, and reasoning, whereas AutoRefine also includes \textit{refinement}, executed immediately after retrieval to synthesise relevant information contained in the documents, and filter out noise \cite{shi2025autorefine}.

Despite the increasing exposure to agent issued queries, no behavioural dataset exists that captures when and what do they search. We introduce a methodology for constructing such datasets and likewise introduce a first dataset, \ourbenchmark, of agent-generated synthetic queries, retrieved documents, and intermediate reasoning steps.
\section{Terminology}

Given an initial query $q_0$, we define an agent's attempt to answer it as an agentic run, or \textit{\arun} for short, i.e., the sequence of actions performed by the agent to answer $q_0$. 
Actions can be operative, such as query formulation, retriever invocation, and answer generation, or reflective, such as reasoning and information refinement, which produce natural-language descriptions that guide subsequent behaviour.
Not all \aruns necessarily yield a final answer: they may terminate prematurely due to run-time errors, or explicit early stop.
Unlike \textit{human search sessions}, which comprise the queries submitted by a user within a time window and may include parallel searches across multiple tabs~\cite{10.1145/1935826.1935875}, \aruns are bound to a single initial query~$q_0$. 

We now define two data abstractions needed to describe our data collection methodology: \textit{frames}, and \textit{traces}.

A frame $f$ records the outputs produced by the \textit{actions} the agent performed during a specific iteration of an \arun:
\begin{align*}
    f=(q, \mathcal{R}_{q}, \mathcal{D})
\end{align*}
where $q$ denotes an intermediate query generated by the agent, $\mathcal{R}_{q}$ is the corresponding ranked list of documents retrieved from the corpus, and $\mathcal{D}$ is the list of descriptions produced by agent-specific reflective operations.

Given an initial query $q_0$, we define as \textit{trace} the ordered collection of frames belonging to the same \arun started with $q_0$, paired with the answer\footnote{The answer can be empty if no answer is returned due to early exit or errors.} $a$ generated by an agent $A$:
\begin{align*}
    \mathcal{T}_A(q_0)= (S,a) \qquad \text{with} \qquad S=(f_0, f_1, \ldots ,f_N),  
\end{align*}
where $f_i$ is the frame associated to the $i$-th iteration of the \arun, composed by a total of $N$ frames.
We refer to $N$ as the \textit{trace length}, and we define \textit{incomplete}, the traces of \aruns with no answer.
While a frame captures  trace is a representation of an \arun capturing all the intermediate and final data an agent produced. 
Frames encode the temporal and causal dependencies between an agent's actions, thus collecting traces enables analysis of action-level interactions and the agent's search session behaviours.

\section{Dataset Construction}\label{sec:methodology}

\subsection{Properties}
We now outline the \textit{intrinsic} and \textit{extrinsic} properties of our dataset, i.e., the characteristics that our dataset should have independently of any use case, and those enabling downstream uses, respectively.

\mypar{Intrinsic}
Each query should be unambiguously linked to its corresponding trace, answer, and all associated frames, and vice versa (\textit{traceability}), with the temporal sequence of actions -- both within and across frames -- fully reconstructable (\textit{chronological order}). 
Each frame should capture all data produced during its iteration, and each trace should include all its constituent frames (\textit{completeness}). Moreover, the \textit{integrity} of agent behaviour must be preserved: logged data should remain raw and unaltered, the agent should not be prompted to decompose or reformulate queries, and no interventions that could influence its decision-making process should be introduced. Finally, the dataset should ensure \textit{diversity} across generator and retriever configurations, query types, corpora, and domains.

\mypar{Extrinsic}
The dataset should support both the optimisation and the rigorous evaluation of retrieval models and systems (\textit{optimisability} and \textit{assessability}), enabling systematic comparison across approaches and configurations. In addition, the methodology should be compatible with a variety of agents and retrieval systems (\textit{interoperability}), allowing different model architectures and implementations to be seamlessly integrated. Finally, both the dataset and the methodology should guarantee \textit{extensibility}, so that new agents, retrievers, tasks, or evaluation protocols can be incorporated over time without requiring structural changes.

\subsection{Methodology}
We now describe our dataset collection methodology, designed to systematically log agentic RAG systems' search behaviours.

\mypar{1. Prompt Construction \& Agentic Run Start}
We build the prompt incorporating the organic query according to each agent's specifics, and we pass it to the generator to start the \arun.

\mypar{2. Iterative Extraction of Frames}
During each \arun execution, all trace data are extracted during the generation’s decoding rather than after the entire sequence is generated --- with each decoding loop corresponding to an iteration. This enables logging incomplete \aruns, ensuring \textit{integrity}.

Since agentic RAG systems coordinate their actions by generating specialised tags, which may vary across agents, we parse the tags using regular expressions, and we intercept\footnote{We could also parse documents via regex, but to reduce storage we log only ids.} retriever calls to record document identifiers. 
While in the remainder of this section we refer to the tags used by Search-R1 and AutoRefine, the approach generalises to other agents with other control tags.
In detail, each time the agent emits a synthetic query within \searchtag tags, we extract it from tags and store it. We then log the identifiers of the retrieved documents by intercepting the retriever's returned results before their content is wrapped within \retrievedtag tags and fed to the generator.

Thoughts, i.e., natural-language descriptions generated during chain-of-thoughts reasoning, are extracted from \thinktag tags.
For agents implementing an explicit refinement\footnote{During \textit{refinement}, the agent extracts and organises information from the retrieved documents.} phase, e.g., \autorefine, refinement step's output is extracted from \refinetag tags~\cite{shi2025autorefine}.
Finally, we extract the final answer from \answertag tags. Answer extraction signals successful \aruns' terminations.

\mypar{3. Run-Time Error Handling}
Data extraction order is model-specific; incorrect ordering may trigger regex failures, so practitioners should follow each model's specifics. 
Models may generate spurious or unexpected text, trigging regex failures; in these cases, we early exit.
By contrast, we consider legitimate, and log, empty retrievals where no documents are retrieved for a query.

\mypar{4. Post-processing \& Storage}
To ensure \textit{integrity}, we do not apply any post-processing or filtering, and we retain incomplete \aruns.
We store traces in a structured format, with each frame and its associated data kept separately to preserve \textit{traceability}, \textit{chronological order}, and selective access. Data formats are extensively described in Section~\ref{sec:storage}.
\section{Dataset}\label{sec:storage} 
\subsection{Format}
\mypar{Iteration Id}
Each frame is assigned an increasing iteration identifier corresponding to its position within an \arun, ensuring \textit{traceability}. For the agents considered, which perform one action per type per iteration, a \textit{chronological order} is inherently preserved.

\mypar{Sharding}
We store each trace in its own directory, with  logged data shared by data type (queries, ranked lists, etc.) and stored in separate \texttt{tsv} files. This design allows users to selectively access only the traces or artifacts relevant to their experiments without downloading the entire dataset.

\mypar{Artifact Formats}
Each trace comprises four \texttt{tsv} artifact files, with rows associated with a \colname{qid} corresponding to the original query:

\begin{itemize}[leftmargin=*]
    \item \textit{answers}:  \colname{qid} (\strtype), \colname{answer} (\strtype)
    \item \textit{synthetic queries}: \colname{qid} (\strtype), \colname{iteration} (\inttype), \colname{llm\_query} (\strtype)
    \item \textit{thoughts}:   \colname{qid} (\strtype), \colname{iteration} (\inttype), \colname{thought} (\strtype)
    \item \textit{ranked lists}:  \colname{qid} (\strtype), \colname{iteration} (\inttype), \colname{docid} (\inttype), \colname{rank} (\inttype)
\end{itemize}
\begin{table}[t!]
    \caption{Statistics and characteristics of the organic query datasets used in the generation of the agentic query traces.}
    \label{tab:input_datasets}
    \centering
    \begin{tabular}{@{}ccccc@{}}
        \toprule
        Dataset & Queries & Query Type & Domain & Query Length \\
        \midrule
         \hotpotqa         & 7,405                    & multi-faced & in-domain  & 15.72$\pm$5.52 \\
         \msmarcodev         & 101,093                  & factoid & out-of-domain     & 6.41$\pm$2.75  \\
         \researchyq     & 96,448                    & researchy & out-of-domain &   7.79$\pm$2.86 \\
         \bottomrule
    \end{tabular}
\end{table}

\subsection{Experimental Setup \& Configurations}
We now describe the experimental setup used to build our dataset.

\mypar{Agents}
We use Search-R1~\cite{jin2025searchr1} and AutoRefine~\cite{shi2025autorefine} as the agents. We exclude R1-Searcher~\cite{song2025r1searcher} as it is outdated, and more advanced agents such as ParallelSearch~\cite{zhao2025parallelsearch} because they are not publicly available yet. Commercial agents are also excluded, as they do not expose the intermediate data required for query tracing.
AutoRefine is only available with the \texttt{Qwen2.5-3B-Base}\footnote{\url{yrshi/AutoRefine-Qwen2.5-3B-Instruct}} generator. 
For Search-R1 we use generators of varying capacities, optimised with exact match and PPO: \texttt{Qwen2.5-3B}\footnote{\url{PeterJinGo/SearchR1-nq_hotpotqa_train-qwen2.5-3b-em-ppo-v0.3}} (\qwensmall), and  \texttt{Qwen2.5-7B}\footnote{\url{PeterJinGo/SearchR1-nq_hotpotqa_train-qwen2.5-7b}} (\qwenmed).
We fix the number of retrieved documents fed to generators to the commonly used $k=3$~\cite{tian2025right}. For the retrieval pipeline, we used two configurations: BM25 only and the MonoElectra cross-encoder~\cite{monoelectra} re-ranking the BM25's top $1000$ results.

\mypar{Input Datasets}
To support \textit{optimisability} and \textit{assessability}, we ground our dataset on IR datasets for which relevance judgements are available. 
To ensure \textit{diversity}, we generated traces for informational queries of varying types: factoid queries from MS MARCO dev (\msmarcodev)~\cite{bajaj2016msmarco}, multi-faced queries from \hotpotqa\footnote{As the agents under study have been fine-tuned on Natural Questions~\cite{kwiatkowski2019nq} and \hotpotqa (train split), we could not generate traces for their queries.} (test split), and for open-ended researchy questions from \researchyq ~\cite{rosset2025researchyquestions}.
We exclude BRIGHT as its tasks are highly domain-specific and knowledge-intensive.
This selection also ensures \textit{diversity} across document types: short web passages (\msmarcodev), Wikipedia articles (\hotpotqa), and long web documents (\researchyq).
We report dataset characteristics and statistics in Table~\ref{tab:input_datasets}.

\mypar{Query Tracing}
During query tracing, we store all frames generated during each \arun, ensuring \textit{completeness}.
When available, agents are provided with document titles  (\hotpotqa, \researchyq). We truncate document text to 512 tokens.
Since re-ranking substantially increases latency, and because our experiments show that it does not affect the dynamics of agentic query traces, we skip it for \researchyq.

\begin{table}[t!]
\caption{Statistics about our \ourbenchmark dataset.}
\label{tab:trace_stats}
\resizebox{\linewidth}{!}{%
\begin{tabular}{@{}ccccccc@{}}
\toprule
\multirow{2}{*}{Generator}   & \multirow{2}{*}{Corpus} & \multirow{2}{*}{Answers}                   & Search                    & Trace & Max. Trace  & Query            \\
   &  &                    &         Calls           &  Length & Length & Length            \\
\midrule 
\multicolumn{7}{c}{\cellcolor{gray!10} BM25 (k=1000) $\gg$ MonoElectra (k=3)}                                                                                                                                                                                               \\ \midrule
\multirow{2}{*}{\qwensmall}  & HQA         & 7.4k                     & 296                            & 0.05$\pm$0.26                       & 4                          & 13.68$\pm$6.93  \\ 
                             & MSM & 100.9k                   & 6k                          & 0.08$\pm$0.38                       & 11                         & 6.88$\pm$2.70   \\
\midrule 
\multirow{2}{*}{\qwenmed}    & HQA        & 7.4k                     & 18.4k                         & 1.36$\pm$1.16                       & 7                          & 8.51$\pm$4.95   \\
                             & MSM          & 100.8k                   & 192.3k                        & 1.13$\pm$1.09                       & 8                          & 6.18$\pm$2.62   \\  
\midrule 
\multirow{2}{*}{\autorefine} & HQA &  5.2k                     & 7.3k                          & 0.71 $\pm$0.84                       & 6                          & 9.36$\pm$4.16   \\
                          & MSM & 82k                    & 56.9k                         & 0.43$\pm$1.17                       & 77                         & 6.36$\pm$2.50   \\
                             \midrule
\multicolumn{7}{c}{\cellcolor{gray!10} BM25 (k=3)}                                                                                                                                                                                                                          \\ \midrule
\multirow{3}{*}{\qwensmall}  & HQA      & 7.4k                     & 284                            & 0.05$\pm$0.35                       & 11                         & 13.55$\pm$6.84  \\
                             & MSM         & 101k                   & 6.6k                          & 0.10$\pm$0.47                       & 17                         & 7.00$\pm$3.00   \\
                             &  RQ           & 96.4k                     & 4.9k                            &  0.52$\pm$6.58                        &  186                          &  7.61$\pm$4.68  \\
\midrule 
\multirow{3}{*}{\qwenmed}    & HQA         & 7.3k                     & 21.7k                         & 1.66$\pm$1.43                       & 9                          & 8.32$\pm$4.52   \\
                             & MSM         & 101k                   & 245k                        & 1.42$\pm$1.27                       & 8                          & 6.18$\pm$2.55   \\
                             & RQ    &   96.3k                   & 294.3k                         & 2.15$\pm$4.30                       & 85                         & 7.47$\pm$2.59   \\
\midrule 
\multirow{3}{*}{\autorefine} & HQA         & 7.4k                     & 7.4k                          & 0.73$\pm$0.87                       & 9                          & 9.18$\pm$4.06   \\
                             & MSM        & 101k                   & 58.2k                         & 0.45$\pm$0.89                       & 54                         & 6.39$\pm$2.49   \\
                             & RQ        & 96.3k                     & 41k                          &  0.32$\pm$0.52                      &    16                       & 7.91$\pm$2.57   \\
  \bottomrule
\end{tabular}
}
\end{table}
\section{Data analysis \& Discussion}
We characterise the differing search behaviours of agents and humans, and explore their dependence on agent capacity and optimisation.

\subsection{Trace Statistics, Workloads \& Query Length}
Table~\ref{tab:trace_stats} summarises the statistics about the traces in \ourbenchmark.

Initially note the significant volume of traces enables model optimisation and  statistical significance testing, supporting the \textit{optimisability} and \textit{assessability} properties.

Additionally, the larger model, \qwenmed, tends to issue a greater number of search calls (by up to $+265\%$) and generate longer traces, substantially increasing workloads, end-to-end inference latency, and resource consumption. 
Trace length variability is high: absolute standard deviation is higher for the larger model, relative standard deviation is higher for smaller ones. Consequently, high-capacity agents yield less predictable IR system workloads, whereas smaller ones, despite generally under-searching, occasionally perform extensive searches or get stuck in long loops. 
Trace length maxima are high (up to 186). According to \citet{tian2025right}, long traces often occur when agents encounter difficulties in converging to high-quality answers.
Unlike humans, who are likely to abandon search sessions when unsatisfied --- typically after $2$ or $3$ query reformulations~\cite{lucchese2013searchsession} --- agents can generate extremely long traces, overburdening IR systems.

Synthetic and organic queries have comparable lengths, except on \hotpotqa, whose queries are multi-faced and in-domain, where agents produced slightly longer queries. Among all the agents, \qwensmall is the one issuing longer queries.

While retrieval effectiveness impacts end-to-end accuracy~\cite{tian2025right}, it does not alter search behaviours. Behavioural differences seem to be governed by agent capacity, optimisation strategy, and test data.

\subsection{Human vs. Agentic Query Reformulations}

\begin{figure*}[t!]
    \centering
    \begin{minipage}{0.21\linewidth}
            \includegraphics[width=\linewidth]{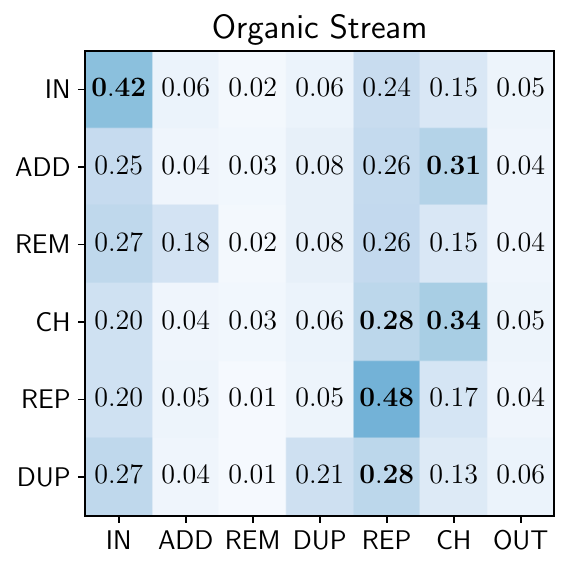}
    \end{minipage}
    \hspace{5mm}
    \begin{minipage}{0.21\linewidth}
            \includegraphics[width=\linewidth]{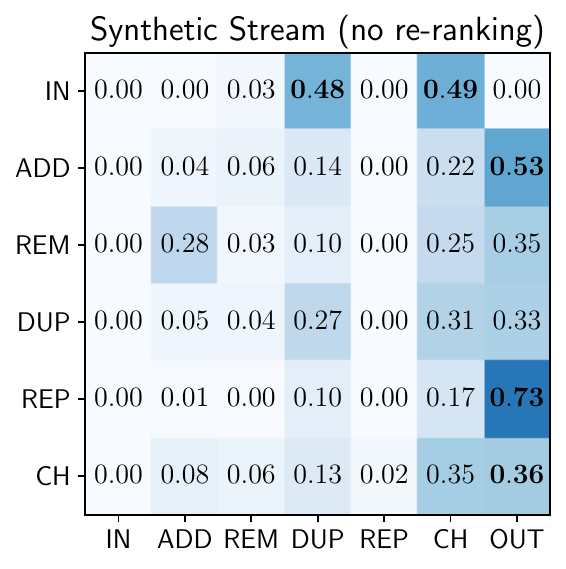}
    \end{minipage}
     \hspace{5mm}
    \begin{minipage}{0.23\linewidth}
            \includegraphics[width=\linewidth]{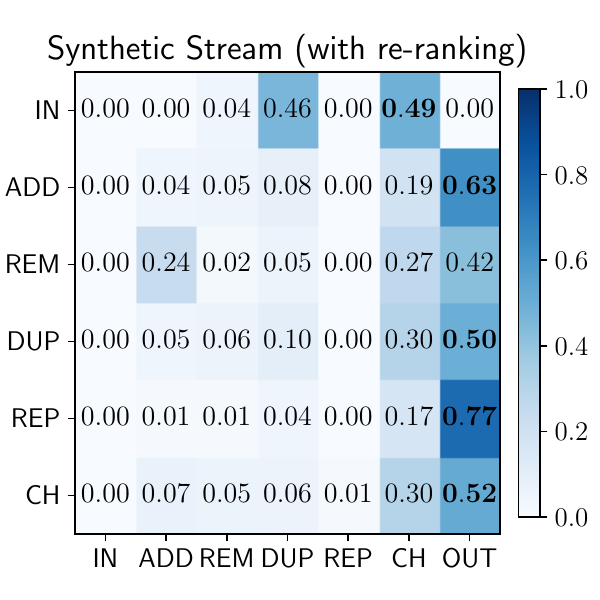}
    \end{minipage}
    \caption{Transition probability matrices representing human (left) and agent search behaviours (centre, right). Rows correspond to the current state and columns to the next state.}
    \label{fig:tm}
\end{figure*}

Following the methodology of~\citet{pass2006aol}, we analyse agentic information seeking behaviours and compare them to those of humans.
We model \aruns as Markov processes.
We map each query $q_i$ in a trace to a state based on its relationship to its preceding queries in the same trace:
\begin{itemize}[noitemsep,nolistsep]
    \item \textsf{IN}: initial state, a new query $q_0$ is issued.
    \item \textsf{ADD}: $q_i$ is an expanded version of $q_{i-1}$.
    \item \textsf{REM}: $q_i$ is a subsequence of $q_{i-1}$.
    \item \textsf{REP}: $q_i$ is a resubmission of $q_{i-1}$.
    \item \textsf{DUP}: $q_{i}$ is an exact duplicate of $q_{j}$ with $j<(i-1)$.
    \item \textsf{CH}: $q_{i}$ results from any other reformulation strategy.
    \item \textsf{OUT}: terminal state (\arun termination).
\end{itemize}
By aggregating all transitions between consecutive queries, for all traces, we build the \textit{transition matrix}. 
Finally, for each non-terminal state $S$, we calculate the termination probability $P(\textsf{OUT} \vert S)$, and insert it into the matrix.
Figure~\ref{fig:tm} shows the mean-aggregated transition matrices  for agents across the different experimental setups.
For comparison, we also show the transition matrix of human reformulations, with values taken from~\cite{pass2006aol}.

Before analysing the matrices, we identify two structural differences.
First, all reformulations in an \arun are tied to the same initial query $q_0$, thus, the \textsf{IN} column is empty (all zeroes). For humans it is non-empty; instead, multiple interleaved searches may collapse into the same session.
Second, while \textsf{REP} in human sessions occurs by clicking on a "show more results" button, likely to increase recall, for \aruns this occurs when $q_i=q_{i-1}$.

We now compare the three matrices in Figure~\ref{fig:tm}, focusing on differences between humans and agents.
For humans the most frequent arrival states are \textsf{IN}, \textsf{REP}, and \textsf{CH}, respectively, highlighting their multi-tasking, results browsing  and substantial reformulations behaviours.
Agents more likely transition to \textsf{OUT}, \textsf{CH}, \textsf{DUP}. The high \textsf{OUT} probability is mainly due to the fact that  \aruns are tied to a single initial query, whereas high \textsf{DUP} probability indicates that agents often regress to previous queries, likely when trapped in long loops. 

The second behavioural difference is that agents favour \textsf{CH} and \textsf{DUP} over \textsf{REP}. Traditional optimisation techniques like caching may be less effective in this new scenario. However, the tendency of agents to re-submit older queries could be exploited to improve retrieval effectiveness and efficiency.

Finally, the transition matrices of the synthetic streams are nearly identical, suggesting that the reformulation behaviour of AI agents is independent from the effectiveness of the retrieval pipeline.

\subsection{Within-Trace Query Evolutions}

\begin{figure*}[t!]
    \centering
    \includegraphics[height=2.7cm,keepaspectratio]{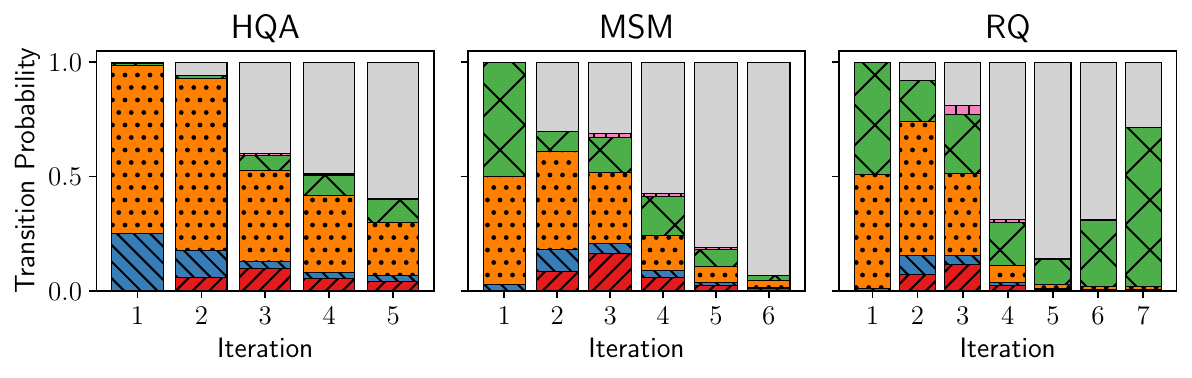}
    \includegraphics[height=2.7cm,keepaspectratio]{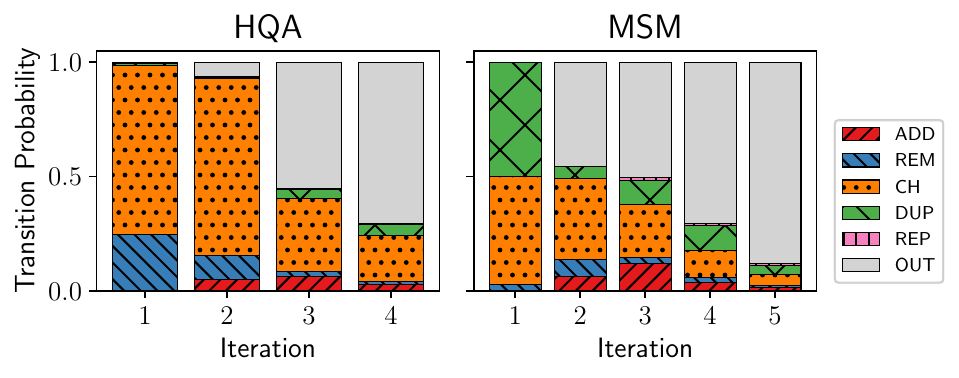}
    \caption{\qwenmed's distribution of transition probabilities across consecutive iterations. Left: retrieval only. Right: retrieval and re-ranking. Stack $i$ shows the transition probabilities from iteration $(i-1)$ to $i$; outlier iterations are omitted.}
    \label{fig:it_tm}
\end{figure*}

We next analyse how reformulation behaviour evolves over the course of \aruns. 
In Figure~\ref{fig:it_tm}, we show the transition probabilities between pairs of consecutive iterations within an agentic run. A comparison with human search sessions would be valuable, but to our knowledge, no public data currently allows such analysis.

Several behavioural patters emerge from Figure~\ref{fig:it_tm}.
First, agents tend to alternate between substantial reformulation (\textsf{CH}) and resubmission of prior queries (\textsf{DUP}). This suggests that they may cycle between trying to differentiate and reverting back to old knowledge upon failure, adopting a trial-and-error approach.

Second, query expansion via \textsf{ADD} and query scope narrowing by removing some terms (\textsf{REM}) are only applied in early iterations, representing initial efforts to refine the organic query.
Subsequent repetitions (\textsf{REP}) emerge from the third iteration onward, likely when agents begin to stall.

Finally, in-domain queries (\hotpotqa) show lower \textsf{DUP} rate compared to out-of-domain ones. This may indicate that the reformulation strategies learnt may generalise poorly and that reverting to prior queries is agent's default when other strategies fail.

Notably, the observed patterns hold regardless of the retriever used. Thus, search behaviours are likely governed by the agent capacity and learned behaviours rather than by retrieval quality.

Overall, our findings suggest a shift in optimisation priorities in agentic IR as compared to traditional IR. Human search queries tend to be mistyped~\cite{sun2012spelling}, short, and ambiguous, thus query expansion~\cite{carpineto2012qexpansion} and other query refinement methods~\cite{ran2025llmgenqueryvariants} are highly effective to improve recall. 
Conversely, in agentic search  they may be less impactful as agents already know how to expand, narrow and change queries, and typically are good writers.

Efficiency-driven optimisations and session-aware diversity enhancement would probably be more beneficial for improving throughput and effectiveness of agentic search traces.
\section{Potential Impact} 

The emergence of agentic search systems necessitates a critical re-examination of the assumptions underlying modern IR infrastructures, which were developed around the characteristics of organic query streams.
A preliminary analysis of a sample of \ourbenchmark's traces reveals several systematic differences between synthetic and organic queries with direct implications for search system design.

\noindent\textit{Query Pre-processing.} Spelling correction and query normalisation techniques may add unnecessary latency when applied indiscriminately to synthetic query streams. Our analysis shows that while approximately 16\% of organic queries contain spelling errors,\footnote{This is based on the English spellchecker built into MacOS 26.2.} consistent with prior query log studies~\cite{sun2012spelling}, only 5\% of synthetic queries exhibit spelling mistakes.

\noindent\textit{Query Understanding.} Intent classification and query understanding components trained on organic query distributions may degrade on synthetic queries due to structural shifts. We observe a 25--50\% reduction in WH-word\footnote{WH-words are interrogative words that typically begin with ``wh'' in English: \textit{what}, \textit{why}, \textit{how}, \textit{when}, \textit{where}, \textit{who}, \textit{which}, and \textit{whose}.} frequency between organic and synthetic queries across all three benchmarks. Additionally, the hapax legomena\footnote{Hapax legomena are words occurring exactly once in a given text.} ratio decreases by 25--40\%, indicating greater term repetition in synthetic queries.

\noindent\textit{Query Processing.} The concentrated term distribution in synthetic queries may improve posting list cache hit rates, but could affect BM25 scoring as repeated terms receive lower IDF weights. This distributional shift may also impact the training of learned sparse models and their query expansion abilities at inference time. Our measurements show synthetic queries exhibit hapax ratios of only 21--43\% compared to 62--78\% in organic streams.

\noindent\textit{Result Caching.} Traditional exact-match caching strategies become less effective for synthetic query streams due to reduced query repetition. Within a single session, human users frequently repeat exact queries (12\% in \researchyq), whereas agents exhibit lower repetition rates (5\% in \researchyq).

\noindent\textit{Semantic Caching.} Semantic caching, where results are shared across semantically similar queries rather than requiring exact matches, presents significant opportunities for synthetic query streams. Queries generated from the same original query share 35\% (\hotpotqa) to 83\% (\researchyq) Jaccard similarity, despite low exact-match overlap.

\ourbenchmark enables the community to further investigate these differences and to develop retrieval systems robust to mixed organic-synthetic query distributions.

\section{Challenges \& Limitations}
Our approach faces several challenges and limitations.
Frequent releases of new agents make \ourbenchmark susceptible to inherent staleness.
Moreover, inference with agentic RAG systems is computationally intensive and requires substantial hardware resources, limiting us to only use a limited number of configurations over a restricted number of query sets.
Since organic, full-scale, production logs are  publicly unavailable, we ground \ourbenchmark on query sets sampled from existing benchmarks. While this prevents the study query frequency and long-tail distributions, it does not hinder the study of how agents \textit{transform} queries during search.
Moreover, the controlled nature of benchmark queries enables reproducible experiments and fair comparisons across agent configurations.
Finally, we do not provide relevance judgments for sub-queries, limiting the ability to evaluate intermediate retrievals. 

\section{Ethics Statement}
\ourbenchmark is derived from publicly available datasets and intended solely for research on agentic search behaviour. The authors do not endorse or assume responsibility for the content or biases in the traces, which do not represent the views of the researchers or their institutions. Users are advised to apply appropriate safety and content filters.

\section{Conclusions \& Future Work}
We release \ourbenchmark, the first dataset designed to support progress in IR for systems operating under agent-driven or mixed human–agent query streams. Grounded on three diverse query sets, \ourbenchmark allows to analyse search behaviours of emerging user segment of agents, and supports the optimisation of retrieval systems and user models around their needs.
We believe \ourbenchmark will also serve as resource for reproducible, offline, and resource-efficient research on agentic search, 
Our experimental analysis highlight several differences between synthetic and organic queries, motivating a re-examination of the assumptions IR systems currently rely on.


\clearpage
\bibliographystyle{template/ACM-Reference-Format}
\balance
\bibliography{bibliography}

@String{CompServ = "Comput. Surveys" }

@inproceedings{jin2025searchr1,
      title = {{Search-R1: Training LLMs to Reason and Leverage Search Engines with Reinforcement Learning}}, 
      author = {Jin, Bowen and Zeng, Hansi and Yue, Zhenrui and Yoon, Jinsung and Arik, Sercan and Wang, Dong and Zamani, Hamed and Han, Jiawei},
      year = {2025},
      booktitle={Proc. COLM},
}

@inproceedings{shi2025autorefine,
    title = {{Search and Refine During Think: Facilitating Knowledge Refinement for Improved Retrieval-Augmented Reasoning}},
    author = {Shi, Yaorui and Li, Sihang and Wu, Chang and Liu, Zhiyuan and Fang, Junfeng and Cai, Hengxing and Zhang, An and Wang, Xiang},
    booktitle = {Proc. NeurIPS},
    year = {2025},
}

@inproceedings{alaofi2022queryvariability,
    author = {Alaofi, Marwah and Gallagher, Luke and Mckay, Dana and Saling, Lauren L. and Sanderson, Mark and Scholer, Falk and Spina, Damiano and White, Ryen W.},
    title = {{Where Do Queries Come From?}},
    year = {2022},
    booktitle = {Proc. SIGIR},
    pages = {2850--2862},
}

@inproceedings{kanoulas2010trec,
    title = {{Session Track Overview}},
    author = {Kanoulas, Evangelos and Carterette, Ben and Clough, Paul and Sanderson, Mark},
    year = {2010},
    booktitle = {Proc. TREC}
}

@inproceedings{bailey2016uqv100,
  title = {{UQV100: A test collection with query variability}},
  author = {Bailey, Peter and Moffat, Alistair and Scholer, Falk and Thomas, Paul},
  booktitle = {Proc. SIGIR},
  pages = {725--728},
  year = {2016}
}

@inproceedings{tian2025right,
  title = {{Am I on the Right Track? What Can Predicted Query Performance Tell Us about the Search Behaviour of Agentic RAG}},
  author = {Tian, Fangzheng and Fang, Jinyuan and Ganguly, Debasis and Meng, Zaiqiao and Macdonald, Craig},
  booktitle = {Proc. IR-RAG Workshop},
  year = {2025},
}

@inproceedings{zhai2025agenticretrieval,
    title = {{Information Retrieval for Artificial General Intelligence: A New Perspective of Information Retrieval Research}},
    author = {Zhai, ChengXiang},
    year = {2025},
    booktitle = {Proc. SIGIR},
    pages = {3876--3886},
}

@inproceedings{sauchuk2022role,
  title = {{On the role of relevance in natural language processing tasks}},
  author = {Sauchuk, Artsiom and Thorne, James and Halevy, Alon and Tonellotto, Nicola and Silvestri, Fabrizio},
  booktitle = {Proc. SIGIR},
  pages = {1785--1789},
  year = {2022}
}

@article{robertson1977probability,
  title={{The probability ranking principle in IR}},
  author={Robertson, Stephen},
  journal={Journal of documentation},
  volume={33},
  number={4},
  pages={294--304},
  year={1977},
}

@inproceedings{zendel2025llmqueries,
    author = {Zendel, Oleg and Al Lawati, Sara Fahad Dawood and Rashidi, Lida and Scholer, Falk and Sanderson, Mark},
    title = {{A Comparative Analysis of Linguistic and Retrieval Diversity in LLM-Generated Search Queries}},
    year = {2025},
    booktitle = {Proc. CIKM},
    pages = {4014--4023},
}

@misc{huang2025surveydeepresearch,
  title = {{Deep research agents: A systematic examination and roadmap}},
  author = {Huang, Yuxuan and Chen, Yihang and Zhang, Haozheng and Li, Kang and Zhou, Huichi and Fang, Meng and Yang, Linyi and Li, Xiaoguang and Shang, Lifeng and Xu, Songcen and others},
  note = {arXiv:2506.18096},
  year = {2025}
}

@inproceedings{askari2023trec2023,
    author = {Askari, Arian and Aliannejadi, Mohammad and Kanoulas, Evangelos and Verberne, Suzan},
    title = {{A Test Collection of Synthetic Documents for Training Rankers: ChatGPT vs. Human Experts}},
    year = {2023},
    booktitle = {Proc. CIKM},
    pages = {5311--5315},
}

@misc{bigdeli2025querygym,
      title = {{QueryGym: A Toolkit for Reproducible LLM-Based Query Reformulation}}, 
      author = {Bigdeli, Amin and Rad, Radin Hamidi  and Incesu, Mert  and Arabzadeh, Negar and Clarke, Charles L. A. and Bagheri, Ebrahim},
      year = {2025},
      note = {arXiv:2511.15996}
}

@inproceedings{rosset2025researchyquestions,
    author = {Rosset, Corbin and Chung, Ho-Lam and Qin, Guanghui and Chau, Ethan and Feng, Zhuo and Awadallah, Ahmed and Neville, Jennifer and Rao, Nikhil},
    title = {{Researchy Questions: A Dataset of Multi-Perspective, Decompositional Questions for Deep Research}},
    year = {2025},
    booktitle = {Proc. SIGIR},
    pages = {3712--3722},
}

@inproceedings{ran2025llmgenqueryvariants,
    author = {Ran, Kun and Alaofi, Marwah and Sanderson, Mark and Spina, Damiano},
    title = {{Two Heads Are Better Than One: Improving Search Effectiveness Through LLM-Generated Query Variants}},
    year = {2025},
    booktitle = {Proc. CHIIR},
    pages = {333--341},
}

@misc{zhao2025parallelsearch,
  title = {{ParallelSearch: Train your LLMs to Decompose Query and Search Sub-queries in Parallel with Reinforcement Learning}}, 
  author = {Zhao, Shu and Yu, Tan and Xu, Anbang and Singh, Japinder and Shukla, Aaditya and Akkiraju, Rama},
  year={2025},
  note={arXiv:2508.09303},
}

@inproceedings{yang2018hotpotqa,
    title = {{H}otpot{QA}: A Dataset for Diverse, Explainable Multi-hop Question Answering},
    author = {Yang, Zhilin  and Qi, Peng  and Zhang, Saizheng  and Bengio, Yoshua  and Cohen, William  and Salakhutdinov, Ruslan  and Manning, Christopher D.},
    booktitle = {Proc. EMNLP},
    year = {2018},
    pages = {2369--2380},
}

@article{kwiatkowski2019nq,
    author = {Kwiatkowski, Tom and Palomaki, Jennimaria and Redfield, Olivia and Collins, Michael and Parikh, Ankur and Alberti, Chris and Epstein, Danielle and Polosukhin, Illia and Devlin, Jacob and Lee, Kenton and Toutanova, Kristina and Jones, Llion and Kelcey, Matthew and Chang, Ming-Wei and Dai, Andrew M. and Uszkoreit, Jakob and Le, Quoc and Petrov, Slav},
    title = {{Natural Questions: A Benchmark for Question Answering Research}},
    journal = {TACL},
    volume = {7},
    pages = {453--466},
    year = {2019},
}

@misc{gao2023retrieval,
  title = {{Retrieval-augmented generation for large language models: A survey}},
  author = {Gao, Yunfan and Xiong, Yun and Gao, Xinyu and Jia, Kangxiang and Pan, Jinliu and Bi, Yuxi and Dai, Yixin and Sun, Jiawei and Wang, Haofen and Wang, Haofen},
  note={arXiv:2312.10997},
  year={2023}
}

@inproceedings{lewis2020rag,
    author = {Lewis, Patrick and Perez, Ethan and Piktus, Aleksandra and Petroni, Fabio and Karpukhin, Vladimir and Goyal, Naman and K\"{u}ttler, Heinrich and Lewis, Mike and Yih, Wen-tau and Rockt\"{a}schel, Tim and Riedel, Sebastian and Kiela, Douwe},
    booktitle = {Proc. NeurIPS},
    pages = {9459--9474},
    title = {{Retrieval-Augmented Generation for Knowledge-Intensive NLP Tasks}},
    year = {2020}
}

@inproceedings{bajaj2016msmarco,
    title = {{MS MARCO: A Human Generated MAchine Reading COmprehension Dataset}},
    author = {Bajaj, Payal and Campos, Daniel and Craswell, Nick and Deng, Li and  Gao, Jianfeng and Liu, Xiaodong and Majumder, Rangan and McNamara, Andrew and Mitra, Bhaskar and Nguyen, Tri and Rosenberg, Mir and Song, Xia and  Stoica, Alina and Tiwary, Saurabh and Wang Tong},
    booktitle = {Proc. InCoCo@NIPS},
    year = {2016},
}

@misc{song2025r1searcher,
    title = {{R1-Searcher: Incentivizing the Search Capability in LLMs via Reinforcement Learning}}, 
    author = {Song, Huatong and Jiang, Jinhao and  Min, Yingqian and  Chen, Jie and  Chen, Zhipeng and Zhao, Wayne Xin  and Fang, Lei and Wen, Ji-Rong},
    year = {2025},
    note = {arXiv:2503.05592},
}

@inproceedings{pass2006aol,
    author = {Pass, Greg and Chowdhury, Abdur and Torgeson, Cayley},
    year = {2006},
    title = {{A picture of search}},
    booktitle = {Proc. InfoScale},
}

@inproceedings{zhang2006behaviour,
  author = {Zhang, Yuye and Moffat, Alistair},
  booktitle = {Proc. ADCS},
  title = {{Some Observations on User Search Behavior}},
  year = {2006},
 pages = {1--8}
    
}

@article{plaat2025surveyreasoning,
    author = {Plaat, Aske and Wong, Annie and Verberne, Suzan and Broekens, Joost and Van Stein, Niki and B\"{a}ck, Thomas},
    title = {{Multi-Step Reasoning with Large Language Models, a Survey}},
    year = {2025},
    volume = {58},
    number = {6},
    journal = CompServ,
}

@inproceedings{wei2022cot,
    author = {Wei, Jason and Wang, Xuezhi and Schuurmans, Dale and Bosma, Maarten and Richter, Brian and Xia, Fei and Chi, Ed and Le, Quoc V and Zhou, Denny},
    booktitle = {Proc. NeurIPS},
    pages = {24824--24837},
    title = {{Chain-of-Thought Prompting Elicits Reasoning in Large Language Models}},
    year = {2022}
}

@inproceedings{shah2023takingsearch,
  author       = {Shah, Chirag  and White, Ryen and Thomas, Paul and Mitra, Bhaskar and Sarkar, Shawon and Belkin, Nicholas J.},
  title        = {{Taking Search to Task}},
  booktitle    = {Proc. CHIIR},
  pages        = {1--13},
  year         = {2023},
}

@article{silvestri2010miningquerylogs,
    author = {Silvestri, Fabrizio},
    title = {{Mining Query Logs: Turning Search Usage Data into Knowledge}},
    journal = {Found. Trends Inf. Retr.},
    year = {2010},
    volume = {4},
    number = {1--2},
    pages = {1--174},
}

@article{jansen2000ipm,
  author       = {Jansen, Bernard and Spink, Amanda and Saracevic, Tefko },
  title        = {{Real life, real users, and real needs: a study and analysis of user queries on the web}},
  journal      = {Inf. Process. Manag.},
  volume       = {36},
  number       = {2},
  pages        = {207--227},
  year         = {2000},
}

@misc{DBLP:journals/corr/abs-2003-13624,
  author       = {Jeffrey Dalton and
                  Chenyan Xiong and
                  Jamie Callan},
  title        = {{TREC} CAsT 2019: The Conversational Assistance Track Overview},
  year         = {2020},
  note       = {arXiv:2003.13624},
}

@inproceedings{DBLP:conf/iclr/YaoZYDSN023,
  author       = {Shunyu Yao and
                  Jeffrey Zhao and
                  Dian Yu and
                  Nan Du and
                  Izhak Shafran and
                  Karthik R. Narasimhan and
                  Yuan Cao},
  title        = {{ReAct: Synergizing Reasoning and Acting in Language Models}},
  booktitle    = {Proc. ICLR},
  year         = {2023},
}

@misc{DBLP:journals/corr/abs-2112-09332,
  author       = {Nakano, Reiichiro  and Hilton, Jacob and Balaji, Suchir  and Wu, Jeff  and Ouyang, Long  and Kim, Christina  and Hesse, Christopher  and Jain, Shantanu  and Kosaraju, Vineet  and Saunders, William  and Jiang, Xu  and Cobbe, Karl  and Eloundou, Tyna  and Krueger, Gretchen  and Button, Kevin  and Knight, Matthew  and Chess, Benjamin  and Schulman, John},
  title        = {{WebGPT: Browser-assisted question-answering with human feedback}},
  year         = {2021},
  note    = {arXiv:2112.09332},
}

@misc{DBLP:journals/corr/abs-2410-09713,
    author       = {Zhang, Weinan and Liao, Junwei and Li, Ning and Du, Kounianhua},
    title        = {{Agentic Information Retrieval}},
    year         = {2024},
    note       = {arXiv:2410.09713},
}

@inproceedings{10.1145/1935826.1935875,
    author = {Lucchese, Claudio and Orlando, Salvatore and Perego, Raffaele and Silvestri, Fabrizio and Tolomei, Gabriele},
    title = {{Identifying task-based sessions in search engine query logs}},
    year = {2011},
    booktitle = {Proc. WSDM},
    pages = {277--286},
}

@article{lucchese2013searchsession,
    author = {Lucchese, Claudio and Orlando, Salvatore and Perego, Raffaele and Silvestri, Fabrizio and Tolomei, Gabriele},
    title = {{Discovering tasks from search engine query logs}},
    year = {2013},
    volume = {31},
    number = {3},
    journal = {ACM Trans. Inf. Syst.},
}

@misc{koneva2025internettraffic,
    title = {{Introducing Large Language Models as the Next Challenging Internet Traffic Source}}, 
    author = {Nataliia Koneva and Alejandro Leonardo García Navarro and Alfonso Sánchez-Macián and José Alberto Hernández and Moshe Zukerman and Óscar González de Dios},
    year = {2025},
    note={arXiv:2504.10688},
}

@inproceedings{wang2025burstgpt,
    author = {Wang, Yuxin and Chen, Yuhan and Li, Zeyu and Kang, Xueze and Fang, Yuchu and Zhou, Yeju and Zheng, Yang and Tang, Zhenheng and He, Xin and Guo, Rui and Wang, Xin and Wang, Qiang and Zhou, Amelie Chi and Chu, Xiaowen},
    title = {{BurstGPT: A Real-World Workload Dataset to Optimize LLM Serving Systems}},
    year = {2025},
    booktitle = {Proc. SIGKDD},
    pages = {5831--5841},
}

@misc{lin2025surveyagents,
    title = {{A Comprehensive Survey on Reinforcement Learning-based Agentic Search: Foundations, Roles, Optimizations, Evaluations, and Applications}}, 
    author = {Lin, Minhua and Wu, Zongyu and Xu, Zhichao and Liu, Hui and Tang, Xianfeng and He, Qi and Aggarwal, Charu and Liu, Hui and Zhang, Xiang and Wang, Suhang},
    year={2025},
    note={arXiv:2510.16724},
}

@inproceedings{wu2025agentic,
    title = {{Agentic Reasoning: A Streamlined Framework for Enhancing {LLM} Reasoning with Agentic Tools}},
    author = {Wu, Junde  and Zhu, Jiayuan  and Liu, Yuyuan  and Xu, Min  and Jin, Yueming},
    booktitle = {Proc. ACL},
    year = {2025},
    pages = {28489--28503},
}

@inproceedings{zhang2025personalisation,
    author = {Zhang, Junhao and Liu, Haiming},
    title = {{Theory-Based User Search Behaviour Modelling and Understanding through Search Log Analysis}},
    year = {2025},
    booktitle = {Proc. CHIIR},
    pages = {298--309},
}

@inproceedings{macavaney2022personalisation,
    author={MacAvaney, Sean and Macdonald, Craig and Ounis, Iadh},
    title =  {{Reproducing Personalised Session Search Over the AOL Query Log}},
    booktitle = {Proc. ECIR},
    year = {2022},
    pages = {627--640},
}

@misc{bacciu2024generatingqueryrecommendationsllms,
      title = {{Generating Query Recommendations via LLMs}}, 
    author = {Bacciu, Andrea and Palumbo, Enrico and Damianou, Andreas and Tonellotto, Nicola and Silvestri, Fabrizio},
      year = {2024},
      note = {arXiv:2405.19749},
}

@inproceedings{zuo2022sessionsearch,
    author = {Zuo, Xiaochen and Dou, Zhicheng and Wen, Ji-Rong},
    title = {{Improving Session Search by Modeling Multi-Granularity Historical Query Change}},
    year = {2022},
    booktitle = {Proc. WSDM},
    pages = {1534--1542},
}

@inproceedings{monoelectra,
    author = {Schlatt, Ferdinand and Fr\"{o}be, Maik and Scells, Harrisen and Zhuang, Shengyao and Koopman, Bevan and Zuccon, Guido and Stein, Benno and Potthast, Martin and Hagen, Matthias},
    title = {{Rank-DistiLLM: Closing the Effectiveness Gap Between Cross-Encoders and LLMs for Passage Re-ranking}},
    year = {2025},
    booktitle = {Proc. ECIR},
    pages = {323--334},
}

@article{walter2025artificial,
  title = {{Artificial influencers and the dead internet theory}},
  author = {Walter, Yoshija},
  journal = {AI \& SOCIETY},
  volume = {40},
  number = {1},
  pages = {239--240},
  year = {2025},
}

@inproceedings{tian2025ragrelevance,
  author       = {Tian, Fangzheng and Ganguly, Debasis and Macdonald, Craig},
  title        = {{Is Relevance Propagated from Retriever to Generator in RAG?}},
  booktitle    = {Proc. ECIR},
  volume       = {15572},
  pages        = {32--48},
  year         = {2025},
}

@inproceedings{sun2012spelling,
    author = {Sun, Xu and Shrivastava, Anshumali and Li, Ping},
    title = {{Fast multi-task learning for query spelling correction}},
    year = {2012},
    booktitle = {Proc. CIKM},
    pages = {285--294},
}

@article{robertson2008ireval,
  author       = {Stephen Robertson},
  title        = {{On the history of evaluation in IR}},
  journal      = {J. Inf. Sci.},
  volume       = {34},
  number       = {4},
  pages        = {439--456},
  year         = {2008},
}

@inproceedings{hawking1999trecweb,
  author       = {Hawking, David  and Voorhees, Ellen M.  and Craswell, Nick  and Bailey, Peter},
  title        = {{Overview of the {TREC-8} Web Track}},
  booktitle    = {Proc. TREC},
  volume       = {500--246},
  year         = {1999},
}

@inproceedings{aliannejadi2024ikat,
  author       = {Aliannejadi, Mohammad  and Abbasiantaeb, Zahra  and Chatterjee, Shubham  and Dalton, Jeffery  and Azzopardi, Leif},
  title        = {{TREC} iKAT 2023: The Interactive Knowledge Assistance Track Overview},
  booktitle    = {Proc. TREC},
  year         = {2023},
}

@article{carpineto2012qexpansion,
    author = {Carpineto, Claudio and Romano, Giovanni},
    title = {{A Survey of Automatic Query Expansion in Information Retrieval}},
    year = {2012},
    volume = {44},
    number = {1},
    journal = CompServ,
}

\end{document}